# Ordered qausi-two-dimensional structure of nanoparticles in semiflexiblering polymer brushes under compression

**Yunfeng Hua, Zhenyu Deng, Yangwei Jiang, Linxi Zhang**[*)]

*Department of Physics, Zhejiang University, Hangzhou, 310027, China*

## Abstract

Molecular Dynamics (MD) simulations are presented for a coarse-grained bead-spring model of ring polymer brushes under compression. Flexible polymer brushes are always disordered during compression, whereas semiflexible brushes tend to be ordered under sufficiently strong compression. Besides, the polymer monomer density of semiflexible polymer brush is very high near the polymer brush surface, inducing a peak value of free energy near the polymer brush surface. Therefore, by compressing nanoparticles (NPs) in semiflexible ring brush system, NPs tend to exhibit a closely packed single layer structure between the brush surface and the impenetrable wall, which provide a new access of designing responsive applications.

[*]Corresponding author. E-mail: lxzhang@zju.edu.cn.

## 1. Introduction

Two-dimensional nanomaterials are an interesting class of materials whose surface area is dominated by one specific crystallographic plane; therefore they are almost crystallographically isotropic. Referenced in literature by a broad range of names such as nanoflakes,[1] nanowalls,[2] nanosheets,[3,4] nanoplates, or nanoplatelets,[5-7] 2D inorganic nanomaterials, not to be confused with 2D carbon-based nanomaterials such as graphene,[8] are typically nanoscale in one dimension yet microscale in the other two. Recently, the utilization of two-dimensional (2D) layered materials has been attracting particular interest because of their unique electronic, structural, and optical properties[9,10], and has promptly become one of the hottest research topics.[11-13] Compared with zero-dimensional (0D) and one-dimensional (1D) materials, 2D-layered materials have several extraordinary advantages,[14] endowing them with promising potential for solar energy harvesting and photocatalytic applications.[15-19] For instance, the 2D-layered materials with high specific surface areas could provide a great number of active sites for various reactions. Additionally, nano-sized 2D-layered materials could act as a support to fabricate various composites with a large interfacial contact.[10,20]

Otherwise, grafted polymeric layers, usually called polymer brushes, are currently used as surface modifiers and have been the subject of intensive investigations for several decades. Due to the delicate interplay between configurational entropy of these polymers, excluded volume, and (solvent mediated) enthalpic interactions, the structure of these soft polymeric layers and their response to external perturbations is characterized by very diverse, complex, and rich properties.[21-23] Despite the fact that polymer brushes are used for various applications,[24] there are still aspects of the complex behavior of such systems that are not yet well understood, and more research still is needed to testify them.

In this paper, we simulate 2D-structure of NPs induced by compressing the NPs in ring polymer brush system. The polymers are adopted to control the effective interactions between NPs and further govern the assembly structures of NPs. Section II gives the model and the simulation method, and Section III describes the results on the density profiles, system pressure, polymer structure, order parameter under various compression degrees, and the ordered structures of NPs. In Section IV, we conclude briefly.

## 2. Model and simulation methods

We employ molecular dynamics simulations based on a coarse-grained model to study the conformations of pure ring polymer brushes and the ordered structure of NPs in ring polymer brushes under compression. Polymer chains are grafted on an impenetrable surface with a grafting density $\sigma_g$, and each chain consists of $N_m$=120 monomers. The number of NPs are fixed to be $N_p$=90. To prevent overlap between any two monomers in the system, a purely repulsive truncated and shifted Lennard-Jones potential is used,

$$U_{LJ}(r_{ij}) = \begin{cases} 4\varepsilon[(\frac{\sigma}{r_{ij}})^{12} - (\frac{\sigma}{r_{ij}})^{6}] + \frac{1}{4} & 0 < r_{ij} < 2^{1/6}\sigma \\ 0 & r_{ij} \geq 2^{1/6}\sigma \end{cases} \quad (1)$$

Where $r_{ij}$ is the spatial distance between monomers, σ is the monomer diameter, and $\varepsilon = k_B T$ ($k_B$ is boltzmann constant and T is temperature of system). Here, the monomer diameter of polymer brushes is σ$_m$=σ and the diameter of NP is fixed to be $\sigma_p = 3\sigma$. We set the mass of the brush monomer to be *m*, and the mass of NP is $(\sigma_p/\sigma)^3$ times of the brush monomer mass.

All bonded monomers in polymer brushes interact with the well-known finitely extensible nonlinear elastic (FENE) potential[25]

$$V_{FENE} = -\frac{KR_0^2}{2}\ln\left[1 - \left(\frac{r}{R_0}\right)^2\right], \quad r < R_0 \quad (2)$$

Where *r* is the distance between two bonded monomers, $K = 30\varepsilon/\sigma^2$ is the spring constant, and $R_0 = 1.5\sigma$ is the maximum distance between bonded monomers.[25]

The stiffness of polymer chain is described by the angle bending potential[26-28]

$$U_b = k_b(1 + \cos\theta) \quad (3)$$

Where $\theta$ is the angle between two consecutive bonds and $k_b$ denotes the bending stiffness.

The polymer brushes are grafted on a planar $30\sigma \times 26\sigma$ impenetrable surface with dense uniform lattice arrangement of grafting sites. The substrate surface exerts a purely repulsive truncated and shifted Lennard-Jones potential described by eqn(1) for all monomers, and the parameters are chosen as $\sigma_{surface} = \sigma$, (the diameter of surface monomer) and $\varepsilon = k_B T$ in eqn(1). Meanwhile, the surface is infinitely extended through periodic boundary conditions in the x-y plane. At the beginning of simulation, the distance between two impenetrable surfaces is $d$=80σ, and NPs are located randomly outside the brush. The grafting surface is immobile and the compressing surface is slowly pressed towards the grafting surface during the compression process. The molecular dynamics simulations are accomplished by performing Langevin dynamics with the open source software LAMMPS[29] under a reduced temperature $T^* = k_B T / \varepsilon = 1.0$. Reduced units of $\varepsilon = 1$, $\sigma = 1$, and $m = 1$ are used, which are chosen to be the units of energy, length, and mass, respectively. The time unit $\tau$ and the friction coefficient $\gamma$ in the molecular dynamics simulations of a Langevin thermostat are set to be $\tau$ =0.001 $\tau_0$ and $\gamma = 1/\tau_0$ ( $\tau_0 = \sqrt{m\sigma^2/\varepsilon}$ is the time unit in our simulation), respectively. A series of simulations with various chain stiffness ($k_b$) and grafting densities ($\sigma_g$) are performed. Data are obtained from about 1×10⁷τ equilibrium time for each case, and total simulation time for each run is about 3.5×10⁷τ. The statistical data mentioned below are averaged over a sufficient number of samples, and the errors of ensemble averages aren't shown in figures because they are less than symbols. Meanwhile, all the simulation snapshots are captured via the Visual Molecular Dynamics (VMD) package. [30]

## 3. Results and discussion

*3.1 Compression-driven conformations of semiflexible ring brushes*

Compression processes of ring polymer brushes are investigated and two typical snapshots of ring polymer brushes with a grafting density of $\sigma_g$ =0.11 are shown in Fig. 1. As indicated, flexible ring polymer brushes with $k_b$ =0 (*a* and *c*) as well as semiflexible ring polymer brushes with $k_b$=160 (b and d, from two different side views) under compression characterized by compression degree $D/h_0$=1.0 (a and b) and $D/h_0$=0.5 (c and d) are displayed. The brush height $h_0$ increases with $k_b$. Here, *D* is the distance between two planar and $h_0$ is the average height of the polymer brushes without compression and defined from when $\Phi(z)$ ( see Fig. 2) has decreased to about 50% of its original value in the flat region of $\Phi(z)$. We set $h_0 \approx 40\sigma$ for flexible ring polymer brush and $h_0 \approx 45\sigma$ for flexible ring polymer. In Figs. 1(a) and 1(c), flexible chains are always disorderly distributed and the density increases gradually as the compression degree increases. In contrast, semiflexible ring chains distribute much more orderly. In the xz-plane, the distribution of ring polymer brush is similar to the semiflexible linear polymer brush shown in our previous work[31]. In the yz-plane, ring-like structures induced by chain stiffness are observed, from which one can infer that the monomer density along *z*-direction near two impenetrable surfaces are greater than that in the central.

The density profiles $\Phi(z)$ of the total monomer density for both flexible ring brushes (Fig. 2a) and semiflexible ring brushes (Fig. 2b) under various compression degrees are shown in Fig. 2. Here, *z* is the distance between brush monomer and the grafting surface along z-direction, and $\Phi(z)$ is the monomer density in the scope from *z* to *z*+Δ*z* (Δ*z* is set to be 0.5σ). In Fig 2a, the monomer density profile for ring polymer brushes and linear polymer brushes[31] are almost identical. For flexible ring polymer brushes with $k_b$=0, compression can lead to a pronounced layering of the monomers near the grafting surface. At the brush surface, the parabolic decay of the density profiles found in the free ring polymer brushes is replaced by the density oscillations in the compressed ring polymer brushes. While for large $k_b$=160, remarkable discrepancies are found. There is no layering of the monomers and the density distribution $\Phi(z)$ near two walls are extremely high which induce a high free-energy near the chain surfaces. The corresponding arrangement of semiflexible ring polymer brushes under compression is shown in the inset figure.

To illustrate the conformations of ring polymer brushes in the process of compression, variations of the angle $\theta_{i,i+1}$ of consecutive bonds along the ring chain backbone at various degrees of compression $D/h_0$ are shown in Fig. 3. Density distributions of the angle $\varphi_{i-1,i,i+1}$ between neighboring bonds $\rho(\varphi)$, the angle $\theta_{i,i+1}$ of consecutive bonds $\rho(\theta)$ at different degrees of compression $D/h_0$ for both linear polymer brush and ring polymer brush are given in Fig. 4. Here, monomers are labeled consecutively $i = 1, 2, \ldots$, starting at the grafting monomer, and $\vartheta_{i,i+1}$ is the angle between $i$-th bond vector and the $z_0$ vector (the unit vector along $z$-direction). In Fig. 3, angle $\theta_{i,i+1}$ is obtained from 5000 samples, each sample consists of 100 polymer chains (here, index $i$ is from $i=1$ to $i=120$, if $i=120$, $\vartheta_{i,i+1}$ represents $\vartheta_{i,1}$). For flexible ring brushes with $k_b=0$, angle $\theta_{i,i+1}$ increases slowly from $i=0$ to $i=50$ and $i=70$ to $i=120$. However, from $i=50$ to $i=70$, $\theta_{i,i+1}$ increasesdrastically. Furthermore, higher compression degree results in the more drastic increase of $\theta_{i,i+1}$. For $k_b=160$, a likely trigonometric function trend is found. For $D/h_0=1.0$, angle $\theta_{i,i+1}$ varies gradually from $\pi/2$ to 0, then to $\pi$, and finally from $\pi$ to $\pi/2$, indicating a circular structure. Comparatively, semiflexible linear chains with $k_b=160$ is displayed in the inset figure, which is completely different from semiflexible ring chain. The polar angles $\theta_{i,i+1}$ for semiflexible linear chain is almost irrelevant to monomer index $i$ when $D/h_0=1.0$. While for $D/h_0=0.5$, $\theta_{i,i+1}$ begins with a gradual increase and finally remains constant.

In Fig. 4, the angle profile between neighboring bonds $\rho(\varphi)$ is hardly varied during compressing for both semiflexible linear brushes and semiflexible ring brushes. For ring polymer brushes, the distribution of the bending energy along the chain backbone is approximately uniform. $\varphi_{i-1,i,i+1}$ is almost constant and $\rho(\varphi)$ is concentrated in a very small range. While for flexible ring polymer brushes, compression makes the density profile of $\varphi_{i-1,i,i+1}$ shift right, and $\varphi_{i-1,i,i+1}$ increases as compression degree increases. The density profile of $\rho(\varphi)$ for semiflexible linear polymer brush is slightly shifted right for stronger compressing. While for flexible ring brush with $k_b=0$, the density profile $\rho(\theta)$ is just like a roller coaster, and the density profile varies from two peaks to just one as $D/h_0$ reaches 0.5. For semiflexible ring brush $k_b=160$, two peak values of the density profile $\rho(\theta)$ becomes greater under stronger compressing, and the corresponding $\theta$ for $D/h_0=0.5$ to the peak value ($\theta\approx1.1$ and $\theta\approx2.0$) is perfectly matched with the Fig. 3.

A particularly useful quantity to describe the conformations of polymer chains is the order parameter, i.e.,

$$S = \frac{1}{2}\left(3<\cos^2\theta'>-1\right) \quad (4)$$

Where $\theta'$ denotes the angle between the middle bond vector $\theta'_{j_{xy}}$ (projection of bond between $i = \frac{N_m}{2} = 60$ and $i' = i+1 = 61$ in xy-plane) of chain and $\theta'_0$, here

$$\theta'_0 = \frac{1}{N_{chain}}\sum_{j=1}^{N_{chain}}\theta'_{j_{xy}} \quad (5)$$

The order parameters for flexible and semiflexible ring brush under various compression degrees are shown in Fig. 5. It is obvious that the order parameter for flexible brush maintains about 0.25. While for semiflexible brush, the order parameter increases as compressing degree increases ($S\approx 0.9$ when $D/h_0=0.5$). Flexible brush is always disordered no matter how the system is compressed, however, for the semeflexible brush, the order degrees exhibits a rapid increase at $D/h_0=1.0$ ($S\approx 0.35$ to $S\approx 0.8$), and finally maintains a stable value. One can conclude that the compression has a great influence on semiflexible brush system, rendering the system more ordered.

In Fig. 6 we show the pressure exerted by a brush of ring polymers with various stiffness as well as linear polymers at height $D$ above the grafting plane. For flexible brushes, the pressure increases gradually during compression, which is in good accordance with the results of K. Binder et. al.[32]. However, the trend for semiflexible brushes is completely different where a drastic increase to a maximum followed by a steady plateau is observed. Under strong compressions, the pressure of the semiflexible brushes is smaller than that of the flexible brushes. Furthermore, the peak value of pressure profile is smaller for semiflexible ring brushes compared to semiflexible linear brushes, while the pressure for ring brushes is greater than linear brushes under strong compressions. The inset shows the schematic illustrations of the compression process. Semiflexible linear brushes are quickly tilting for stronger compressing. For semiflexible ring brushes, deformation occurs for small compression degree ($D/h_0>0.92$), and titling is followed as compressing degree increases.

### 3.2 Ordered structures for Nanoparticles in semiflexible ring polymer brushes

Due to the nontrivial monomer density distribution of semiflexible ring polymer brushes, one may obtain ordered structures of NPs by compressing NPs on the brush surface. Fig. 7 shows the conformations of nanoparticles (NPs) in ring polymer brushes at various compression degrees $D/h_0$=1.0 (a and b) and $D/h_0$=0.5 (c and d). For both flexible and semiflexible ring brushes, all NPs are accumulated near the polymer brush surface during compression. For flexible ring polymer brushes, NPs are randomly located at the brush surface, owing to the uniform distribution of the polymer ring brush and the gradual increase of the polymer brush density. While for semiflexible ring polymer brushes, since the monomer density is relatively higher near the polymer surface, NPs are closely packed in one single layer. Besides, for larger compression degree, NPs are more orderly packed. The quasi 2D ordered structure of NPs is of great importance in various applications as it improves scratch resistance, exposes catalytic components for bioreactions and inhibits dewetting from low energy substrate.

To analysis the distribution of NPs in more detail, we plot the density profiles of NPs, $\rho_n(z)$, as a function of $z$ for linear brush and ring polymer brush at three compression degrees $D/h_0$=1.0, 0.75, and 0.5, and the results are shown in Fig. 8. We set the value $h_0 \approx 58\sigma$ for semiflexible linear brush with $N_m$=60 and $\sigma_g$=0.22. The compressing surface is shown in Fig. 8 and the arrowhead means that the system is compressed from right to left. The result for semeflexible linear polymer brushes is in good agreement with our previous work[31]. When the surface is compressed to $D/h_0$=1.0, NPs are penetrated into the brush and uniformly distributed in the brush. While at a large compression degree of $D/h_0$=0.5, NPs are aggregated near the two impenetrable surfaces. For ring brushes, the density profile is in good agreement with the snapshots shown in Fig. 7. All NPs are located between the polymer brush surface and compressing surface. As the compression degree increases, NPs are packed in a single layer for semiflexible ring brushes. While for flexible ring polymer brushes, NPs are randomly located near the brush surface, irrelevant to the compression degree.

In Fig. 9, we calculate the radial distribution function (RDFs) between NPs in the single layer near the compressing surface. NPs pack more closely and more orderly for $k_b$=160, and the largest pick value for $g(r_{xy})$ is much greater than that of $k_b$=0. Besides, as shown in the inset figure, since the diameter of the NP is $\sigma_p$=3$\sigma$, if NPs are closely arranged, the peak values should occur at $r_{xy}$=3.0, 5.2, 6.0, and 7.9. The correctly corresponding value with the peak of $g(r_{xy})$ profile appear at $r_{xy}$=3.05, 5.35, 6.15, and 8.15, which further indicates a highly ordered structure of NPs for $k_b$=160.

Finally, the free energy for a single NP in ring semiflexible polymer brushes is calculated via Umbrella sampling method[33] as shown in Fig. 10. The unit of free energy is $k_BT$. The minimum free energy of the system is located near the compressing surface, which explains why NPs prefer to stay near the compressing surface. Therefore, the aggregation of NPs can be well understood through free energy landscape.

## 4. Conclusion

Extensive simulations of flexible as well as semiflexible ring polymer brushes under compression have been presented. Flexible ring polymer brushes are disordered while semiflexible ring polymer brushes tend to tilt and exhibit an ordered structure under sufficient compression degree. Meanwhile, the monomer density near ring polymer brush is extremely high, which yields a peak value of free energy. Therefore, NPs can't penetrate into ring polymer brush and form a quasi 2D ordered structure near the brush surface under strong compression, which proves a new access of designing quasi 2D materials.

**Acknowledgement**

This research was financially supported by the National Natural Science Foundation of China (Grant Nos. 21374102, 21674082, 21674096).

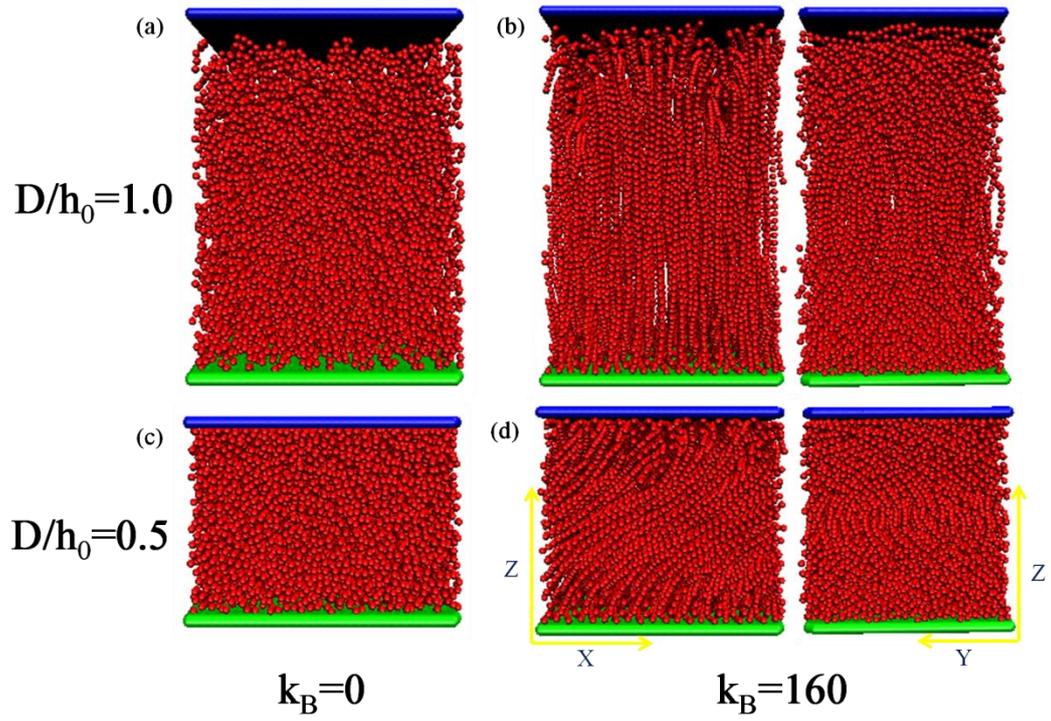

Fig. 1. Snapshots of ring polymer brushes for the case of $\sigma_g$=0.11. Two choices of $k_b$, $k_b$=0 (a and c) and $k_b$=160 (b and d, from two different side views) under compression characterized by $D/h_0$=1.0 (a and b) and $D/h_0$=0.5 (c and d) are distributed.

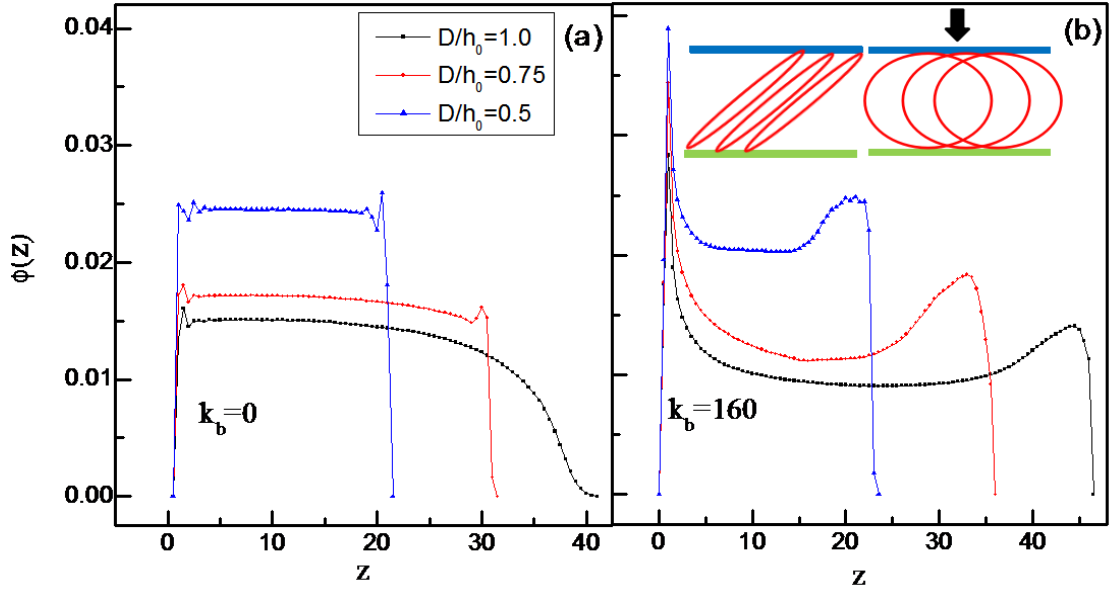

Fig. 2. Density distribution of the effective monomers in the brush, $\Phi(z)$, plotted vs. distance $z$ from the grafting surface, for three choices of $D/h_0$, namely $D/h_0$=1.0, $D/h_0$=0.75 and $D/h_0$=0.5. Two values of $k_b$ are included: (a) flexible ring polymer ($k_b$=0), and (b) semiflexible ring polymer ($k_b$=160). Inset figure shows the corresponding arrangement of semiflexible brushes in $xy$-plane. Here $N_m$=120, and $\sigma_g$=0.11

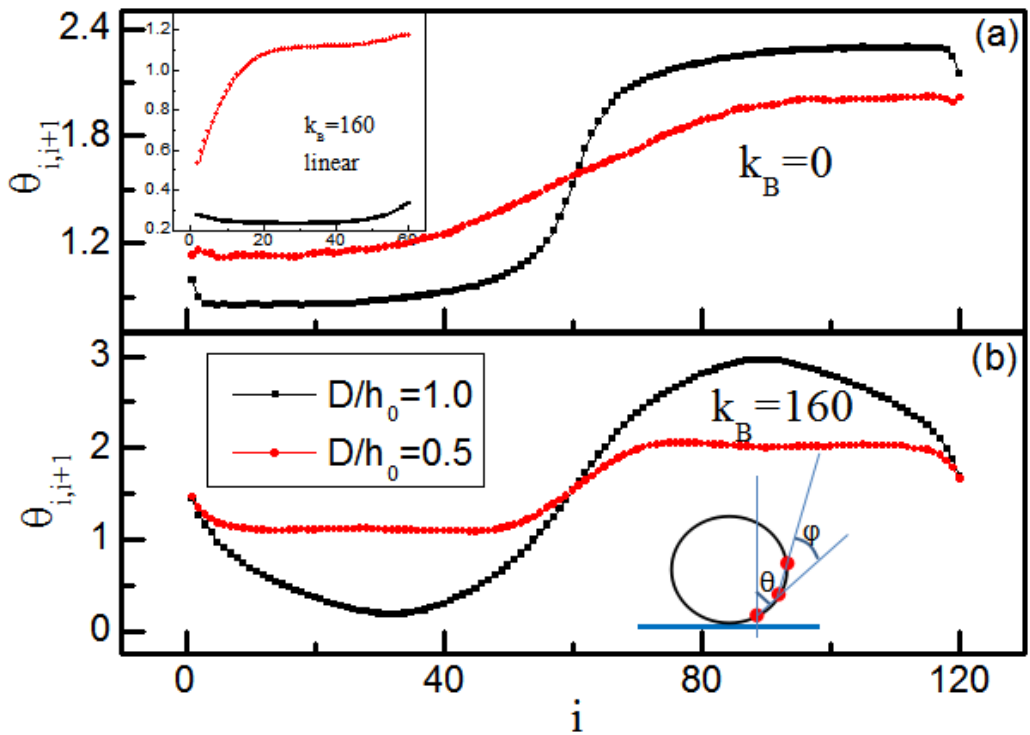

Fig. 3  Variation of the angle θ$_{i,i+1}$ between two consecutive bonds along the ring chain backbone as well as the normal to the grafting surface under different degrees of compression $D/h_0$. Two values of $k_b$ are included: (a) flexible ring polymer ($k_b$=0), and (b) semiflexible ring polymer ($k_b$=160), and the inset shows the case of semiflexible linear chain for $k_b$=160.

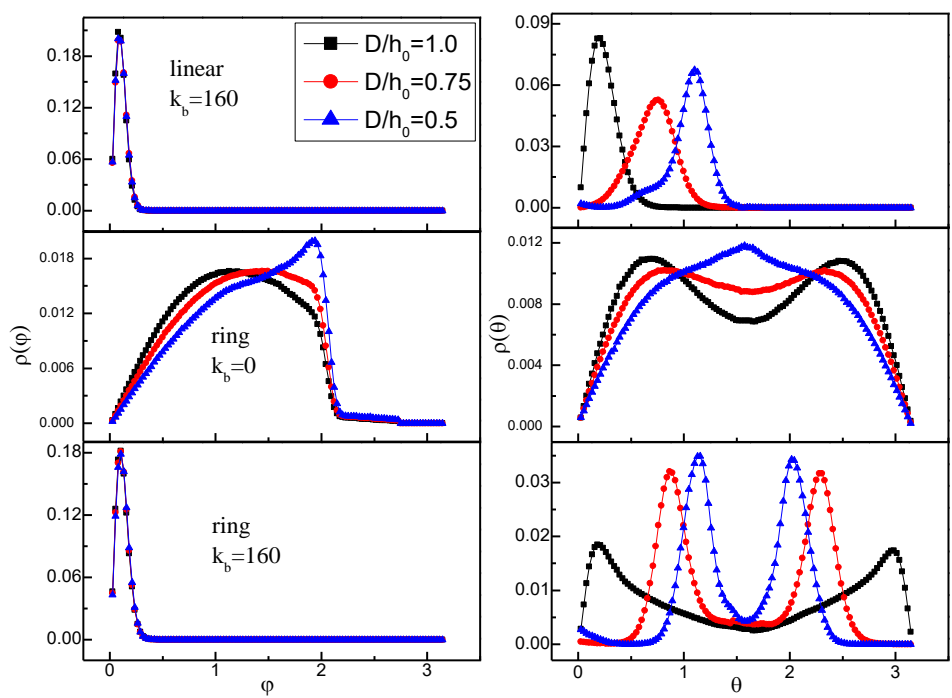

Fig. 4. Density distribution of the angle $\varphi_{i-1, i, i+1}$ between neighboring bonds $\rho(\varphi)$ and the angle $\theta_{i,i+1}$ of consecutive bonds $\rho(\theta)$ at different degrees of compression $D/h_0$ for linear and ring chains.

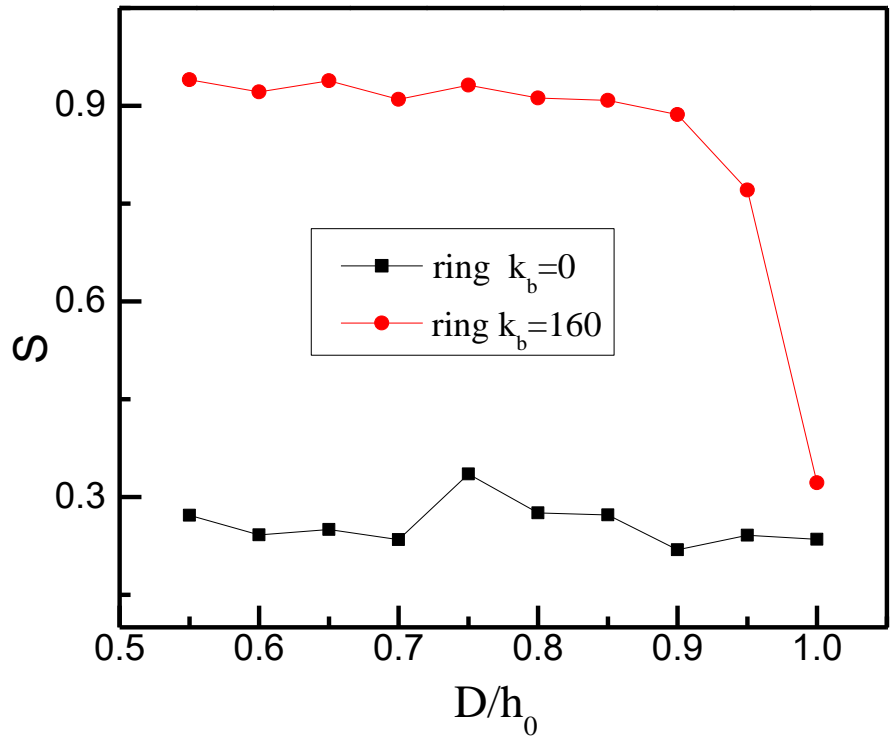

Fig. 5. The order-parameter *S* of the orientation of the bond between 59$^{th}$ and 60$^{th}$ monomers in polymer brushes (represented by unit vectors in the xy-plane), plotted vs. *D/h$_0$* for flexible and semiflexible ring brushes. Here N$_m$=120, and *σ$_g$*=0.11.

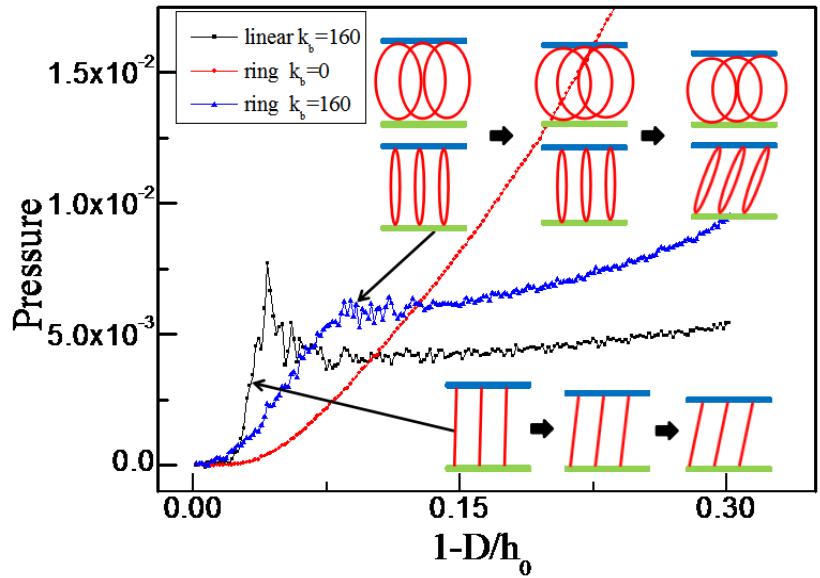

Fig. 6. Pressure P along z-direction for ring brush with chain length $N_m$=120, grafting density $\sigma_g$=0.11 (both flexible brush and semiflexible brush are included) and linear brush with $N_m$=60, $\sigma_g$=0.22 plotted vs. the degree of compression, 1-$D/h_0$. Schematic displays the transformation of polymer brushes in the process of compression.

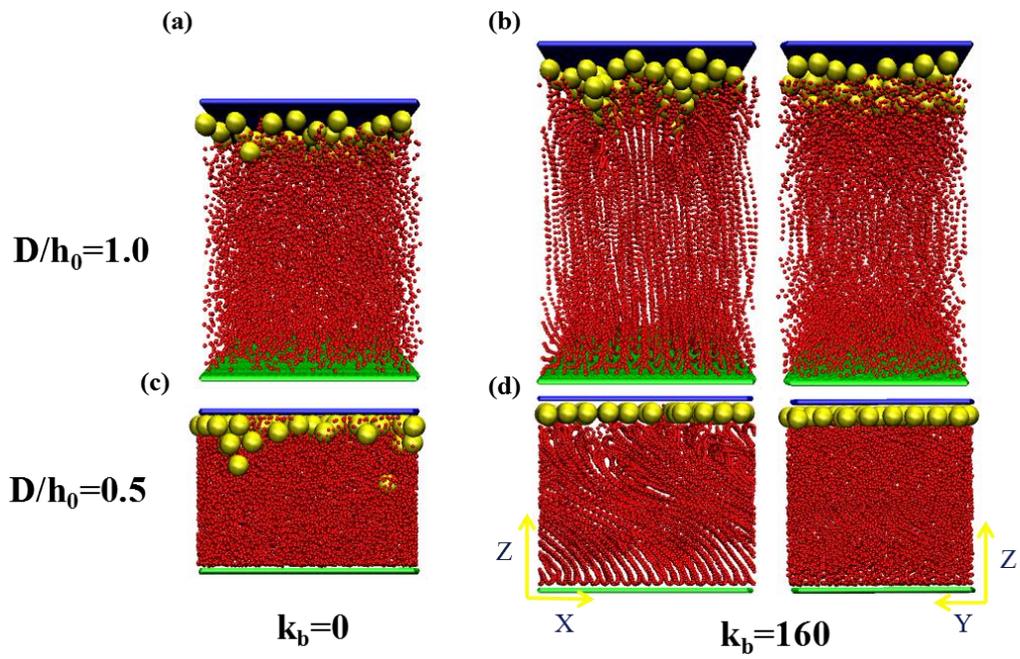

Fig. 7. Snapshots showing the conformations of nanoparticles (NPs) in ring polymer brushes for the case of $N_m$=120, $\sigma_g$=0.11, and two choices of $k_b$, $k_b$=0 (a and c) and $k_b$=160 (b and d, from two different side views).

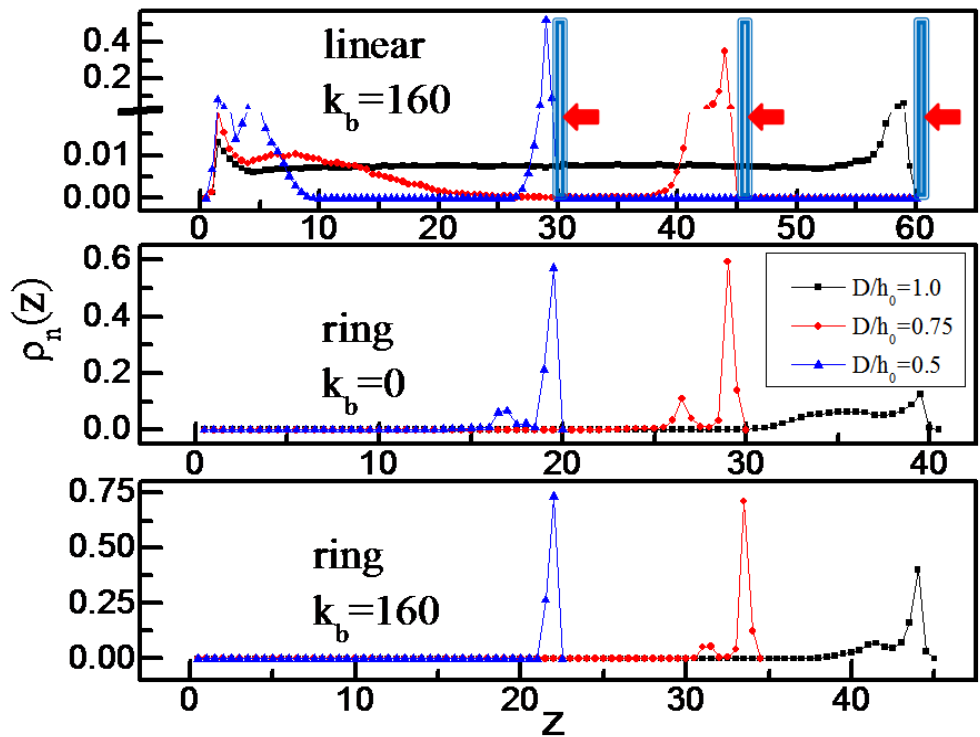

Fig. 8. Density distributions of NPs $\rho_n(z)$ as a function of *z* in linear brush and ring polymer brushes, as indicated, at three compression degrees $D/h_0$=1.0, 0.75, and 0.5.

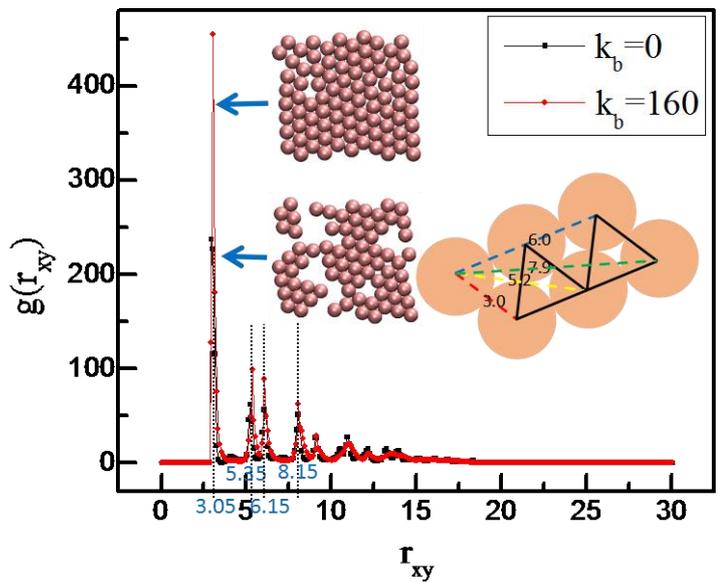

Fig. 9  Pair distribution functions $g(r_{xy})$ between NPs in the first layer close to the compressing surface plotted vs. the distance between NPs $r_{xy}$. Inset shows the arrangement of the NPs in the first layer, and the schematic displays the distances corresponding to the peak values of $g(r_{xy})$.

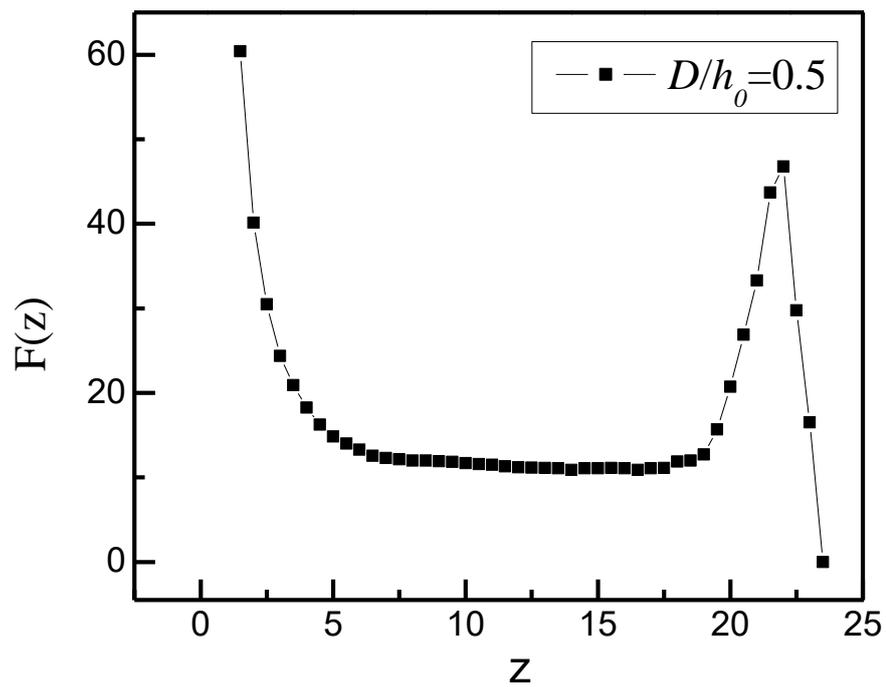

Fig. 10. Free energy F(z) for one NP in semiflexible ring polymer brushes with $k_b$=160 at a compression degree of $D/h_0$=0.5.